%% file: EPring_8.tex
\documentclass[aps,showpacs,twocolumn]{revtex4-2}
\usepackage[utf8]{inputenc}
\usepackage{amsfonts}
\usepackage{amssymb}
\usepackage{amsmath}
\usepackage{graphicx}
\usepackage{epsfig}
\usepackage{subfigure}
\usepackage{appendix}
\usepackage{hyperref}
\usepackage{color}

\hypersetup{hypertex=true, colorlinks=true, linkcolor=blue, urlcolor=blue, citecolor=blue}

\begin{document}

\title{Exceptional nodal rings emerging in spinful Rice-Mele chains}
\author{E. S. Ma and Z. Song}
\email{songtc@nankai.edu.cn}
\affiliation{School of Physics, Nankai University, Tianjin 300071, China}
\begin{abstract}
The Weyl exceptional nodal lines usually occur in 3D topological semimetals,
but also emerge in the parameter space of 1D systems. In this work, we study
the impact of dissipation on the nodal ring in a 3D topological semimetal.
We find that the energy spectrum becomes fully complex in the presence of
dissipation, and the original nodal ring is split into two exceptional
rings. We introduce a vortex field in the momentum space, which is generated
from the spectrum, to characterize the topology of the exceptional rings.
This provides a clear physical picture of the topological structure. The two
exceptional rings act as two vortex filaments of a free vortex flow with
opposite circulations. In this context, the 3D topological semimetal is the
boundary separating two quantum phases identified by two configurations of
exceptional rings. We also propose a 1D model that has the same topological
feature in the parameter space. It provides a simple way to measure the
topological invariant in a low-dimensional system. Numerical simulations
indicate that the topological invariant is robust under the random
perturbations of the system parameters.
\end{abstract}

\maketitle

\section{Introduction}

The search for new quantum phases of matter is a central theme in modern
condensed-matter physics. The emergence of topological materials counts
among the most fascinating discoveries in the field. Topological quantum
states have attracted immense attention in recent years since they produce
fundamentally new physical phenomena and have potential applications in
novel devices.\ Their unique electronic structure hosts exotic phases such
as topological insulators and Weyl semimetals. Research on these topics has
rapidly evolved from theoretical proposals to material synthesis and
characterization. One particularly intriguing direction is the realization
of topological nodal-line semimetals (NLSMs) \cite{burkov2011}, in which the
conduction and valence bands touch along continuous lines or loops within
the three-dimensional Brillouin zone (BZ). In parallel, it has been proposed
that there exists Weyl exceptional ring in a non-Hermitian system, which is
characterized by both a quantized Chern number and a quantized Berry phase 
\cite{xuweyl2017}. On the other hand, the objects of study have expanded
beyond electronic systems to include non-electronic counterparts such as
mechanic \cite{kane2014topological,huber2016topological}, acoustic \cite%
{yang2015topological,zhang2018topological}, photonic \cite%
{Haldane2008possible,lu2014nat,ozawa2019topological}, and ultracold atomic
systems \cite%
{atala2013direct,goldman2016topological,wang2013topological,nakajima2016,lohse2016thouless}%
.

A genuine topological insulator or semimetal is at least two dimensional,
which supports topological invariants, such as Chern number and quantized
Berry flux. However, topological characterization can be demonstrated in
one-dimensional systems. This allows the simplification of topology
detection. A prototype example is the Rice-Mele (RM) model \cite%
{rice1982elementary}. It is well known that adiabatic charge pumping in the
RM model is quantized when the system parameters vary slowly with time along
a loop lines via a chain system \cite%
{wang2018dynamical,wang2019robustness,wang2018rm,ma2024topological,xiao2010berry,asboth2016short,vanderbilt2018berry}%
.%

In this work, we study a non-Hermitian variant of a 3D topological semimetal
and its connection to 1D systems. We find that the energy spectrum of a 3D
topological semimetal becomes fully complex and the original nodal ring is
split into two exceptional rings in the presence of dissipation. Unlike the
nodal loop in the original Hermitian system, where the topological features
are characterized by a vector field based on the Berry curvature of the
energy band, the topological features of the exceptional nodal loop are
characterized by a vector field based on the complex spectrum. Furthermore,
we show that the exceptional rings in a 3D topological semimetal correspond
to vortex filaments of a free vortex flow. In addition, we propose a 1D RM
ladder as a low-dimensional counterpart of the 3D model. This allows us to
establish an experimental scheme to measure the winding number. This scheme
provides a simple way since only measurements of single energy levels are
needed, rather than those of a whole energy band. Numerical simulations
indicate that the topological invariant is robust under random perturbations
of the system parameters, which break the translational symmetry.

The rest of the paper is organized as follows. In Sec. \ref{Hamitlonian and
solution}, we begin by introducing the Hamiltonian, its symmetries, and
solutions. In Sec. \ref{Field induced by EP rings}, we demonstrate that the
exceptional rings in a 3D topological semimetal correspond to a vortex field
in momentum space, which is generated from the complex spectrum. On the
basis of these results, in Sec. \ref{Rice-Mele chain}, we propose a 1D RM
model that has the same topological features in the parameter space. This
model provides a simple way to measure topological invariants in a
low-dimensional system. Following these formulations, in Sec. \ref%
{Robustness to perturbations}, numerical calculations are used to compare
the results obtained from a perfect system with those from a system with
disorder perturbations. Finally, Sec. \ref{Summary} summarizes our findings.
Some detailed derivations are given in the Appendix.

\section{Model and exceptional rings}

\label{model and exceptional ring}

\begin{figure*}[tbh]
\centering
\includegraphics[width=1.0\textwidth]{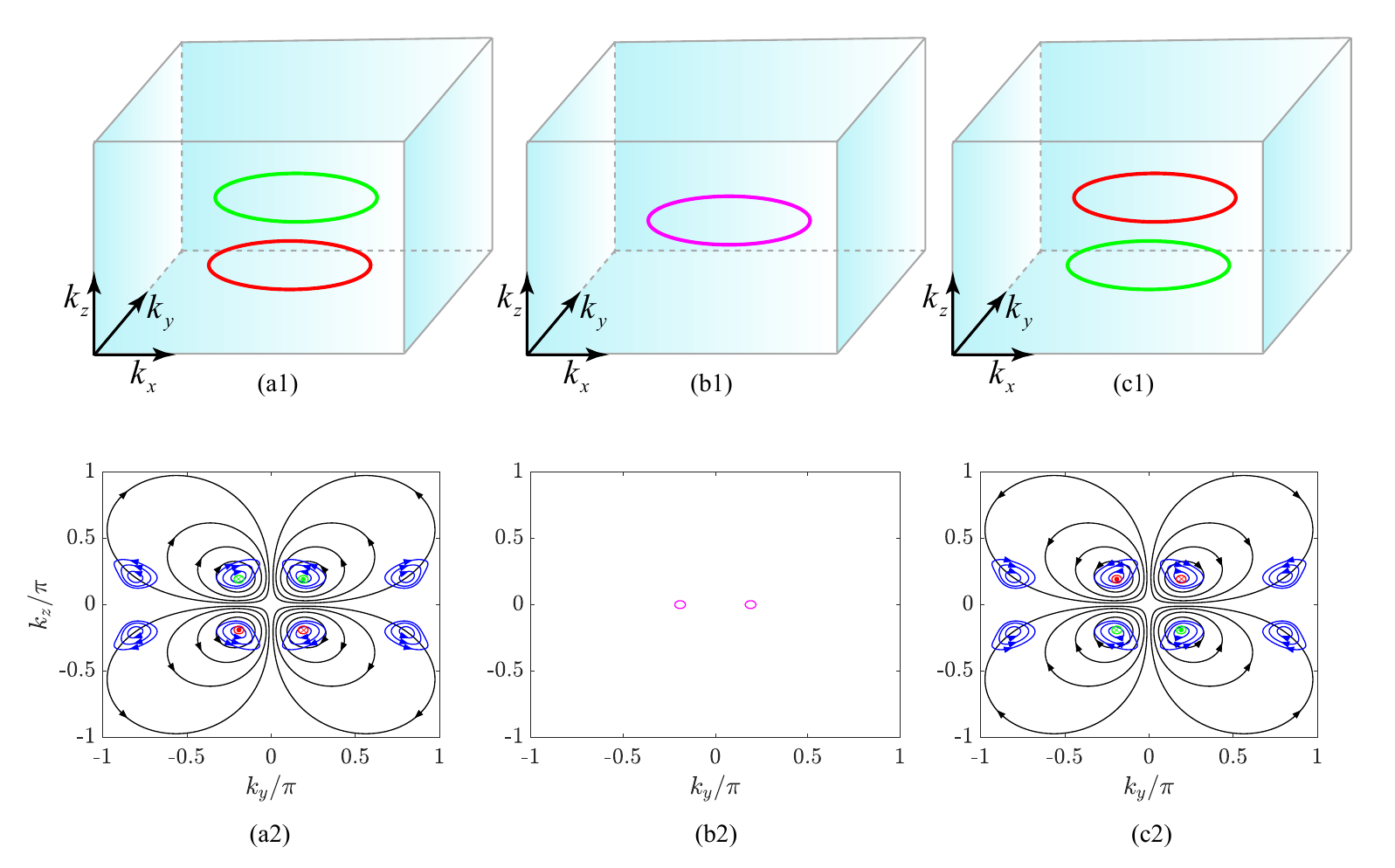}
\caption{Schematic of the exceptional rings described by Eq. (\protect\ref%
{semimetalER}) in three-dimensional momentum space, together with the
gradient field of the eignenenrgy argument defined in the Eq. (\protect\ref%
{field}) and the analogous field of the RM modal under the parameter
transformation of Eq. (\protect\ref{Ptransformation}). Panels (a1), (b1) and
(c1) correspond to $\protect\alpha=0.6$ and $\protect\beta =0.6$, $0$, and $%
-0.6$, respectively. Panels (a2), (b2) and (c2) show the corresponding field
lines of $\mathbf{P}$ (black) and the approximate field lines of the RM
model (blue) near the exceptional points in the $k_{y}-k_{z}$ plane; arrows
indicate the directions. For the semimetal model, the field $\mathbf{P}$
forms a free vortex flow whose circulation follows the right-hand screw
rule, marked by green or red circles.}
\label{fig1}
\end{figure*}

Our starting point is a non-Hermitian system derived from a
three-dimensional Hermitian model of a nodal-ring semimetal. The Hamiltonian
in the momentum space has the following form 
\begin{equation}
H\left( \mathbf{k}\right) =k_{x}s_{x}+k_{y}\tau _{y}s_{y}+k_{z}s_{z}+m\tau
_{x}s_{x},
\end{equation}%
where $\tau _{i}$ and $s_{i}$ ($i=x,y,z$) are Pauli matrices for two isospin
degrees of freedom. The explicit matrix form of the Hamiltonian is 
\begin{equation}
H\left( \mathbf{k}\right) =\left( 
\begin{array}{llll}
k_{z} & k_{x} & 0 & -k_{y}+m \\ 
k_{x} & -k_{z} & k_{y}+m & 0 \\ 
0 & k_{y}+m & k_{z} & k_{x} \\ 
-k_{y}+m & 0 & k_{x} & -k_{z}%
\end{array}%
\right) ,  \label{Hk}
\end{equation}%
which has been systematically studied in Ref. \cite{fang2015} in the case
with real $m$. In this work, we extend the model to a non-Hermitian one by
taking complex $m $, i.e., 
\begin{equation}
m=\alpha +i\beta .
\end{equation}%
No matter $m$\ is real or complex, $H\left( \mathbf{k}\right) $ is chirally
symmertric, because 
\begin{equation}
\Lambda H\left( \mathbf{k}\right) \Lambda ^{-1}=-H\left( \mathbf{k}\right) ,
\label{chiral H}
\end{equation}%
where the matrix $\Lambda $\ is 
\begin{equation}
\Lambda =\left( 
\begin{array}{llll}
0 & 0 & 0 & i \\ 
0 & 0 & -i & 0 \\ 
0 & i & 0 & 0 \\ 
-i & 0 & 0 & 0%
\end{array}%
\right) .
\end{equation}%
The eigen energy of $H\left( \mathbf{k}\right) $ is 
\begin{equation}
\varepsilon _{\mu ,\nu }=\mu \sqrt{k_{z}^{2}+\left( \sqrt{k_{x}^{2}+k_{y}^{2}%
}+\nu m\right) ^{2}},
\end{equation}%
with $\mu $, $\nu =\pm $, in associated with the corresponding eigenstate 
\begin{equation}
\left\vert \psi _{\mu ,\nu }\right\rangle =\left( 
\begin{array}{l}
k_{x}^{2}+\left( k_{y}-m\right) \left( k_{y}-\nu \sqrt{k_{x}^{2}+k_{y}^{2}}%
\right) \\ 
k_{x}\left( \varepsilon _{\mu ,\nu }-k_{z}\right) \\ 
k_{x}\left( m^{2}-k_{y}^{2}-\xi \right) /2m \\ 
\left( k_{z}-\varepsilon _{\mu ,\nu }\right) \left[ \left( k_{y}+m\right)
^{2}+\xi \right] /2m%
\end{array}%
\right) ,
\end{equation}%
where $\xi =k_{x}^{2}+k_{z}^{2}-\varepsilon _{\mu ,\nu }^{2}$. Obviously the
nodal lines occur at $\mathbf{k}=\mathbf{k}_{\text{\textrm{c}}}$, which is
determined by the equation $\varepsilon _{\mu ,+}\left( \mathbf{k}_{\text{%
\textrm{c}}}\right) =0$ or $\varepsilon _{\mu ,-}\left( \mathbf{k}_{\text{%
\textrm{c}}}\right) =0$. Here we only consider the case with $\alpha
\geqslant 0$ for simplicity. Similar conclusions can be obtained for $\alpha
<0$. Then the nodal line obeys the equations 
\begin{equation}
\sqrt{k_{\text{\textrm{c}}x}^{2}+k_{\text{\textrm{c}}y}^{2}}=\alpha ,k_{%
\text{\textrm{c}}z}=\pm \beta ,  \label{semimetalER}
\end{equation}%
which represent a pair of concentric circles of radius $\alpha $ in
three-dimensional momentum space. The distance between these two circles is $%
2\left\vert \beta \right\vert $. Subsequently, we have $\xi _{\text{\textrm{c%
}}}=k_{x}^2+\beta^2$. The zero-energy eigenstates have the form

\begin{equation}
\left\vert \psi _{\mu ,-}\left( \mathbf{k}_{\text{\textrm{c}}}\right)
\right\rangle =\left( 
\begin{array}{l}
-i\left( \alpha +k_{\mathrm{c}y}\right) \\ 
-k_{\mathrm{c}x}k_{\mathrm{c}z}/\beta \\ 
ik_{\mathrm{c}x} \\ 
\left( \alpha +k_{\mathrm{c}y}\right) k_{\mathrm{c}z}/\beta%
\end{array}%
\right) ,
\end{equation}%
\textbf{\ }and we have
\begin{equation}
\left\vert \psi _{+,-}\left( \mathbf{k}_{\text{\textrm{c}}}\right)
\right\rangle =\left\vert \psi _{-,-}\left( \mathbf{k}_{\text{\textrm{c}}%
}\right) \right\rangle ,
\end{equation}%
which is the coalescing state. However, for $\beta =0,\varepsilon _{\mu ,\nu
}=0$\ gives
\begin{equation}
\sqrt{k_{\text{\textrm{c}}x}^{2}+k_{\text{\textrm{c}}y}^{2}}=\alpha ,k_{%
\text{\textrm{c}}z}=0,
\end{equation}%
and the corresponding zero-energy eigenstates can be expressed as 
\begin{eqnarray}
\left\vert \psi _{+,-}\left( \mathbf{k}_{\text{\textrm{c}}}\right)
\right\rangle &=&\left( 
\begin{array}{l}
k_{\mathrm{c}y}+\alpha \\ 
0 \\ 
-k_{\mathrm{c}x} \\ 
0%
\end{array}%
\right) , \\
\left\vert \psi _{-,-}\left( \mathbf{k}_{\text{\textrm{c}}}\right)
\right\rangle &=&\left( 
\begin{array}{l}
0 \\ 
-k_{\mathrm{c}x} \\ 
0 \\ 
k_{\mathrm{c}y}+\alpha%
\end{array}%
\right) ,
\end{eqnarray}%
we have 
\begin{equation}
\left\langle \psi _{+,-}\left( \mathbf{k}_{\text{\textrm{c}}}\right) |\psi
_{-,-}\left( \mathbf{k}_{\text{\textrm{c}}}\right) \right\rangle =0.
\end{equation}%
which indicate that the nodal ring is split into two exceptional rings by
nonzero $\beta $. In Fig. \ref{fig1} (a1), (b1), and (c1) we present the
schematic diagram for the exceptional rings at different $\beta $. In the
following, we investigate the topological characters of the exceptional
rings.

\begin{figure}[tbh]
\centering
\includegraphics[width=0.5\textwidth]{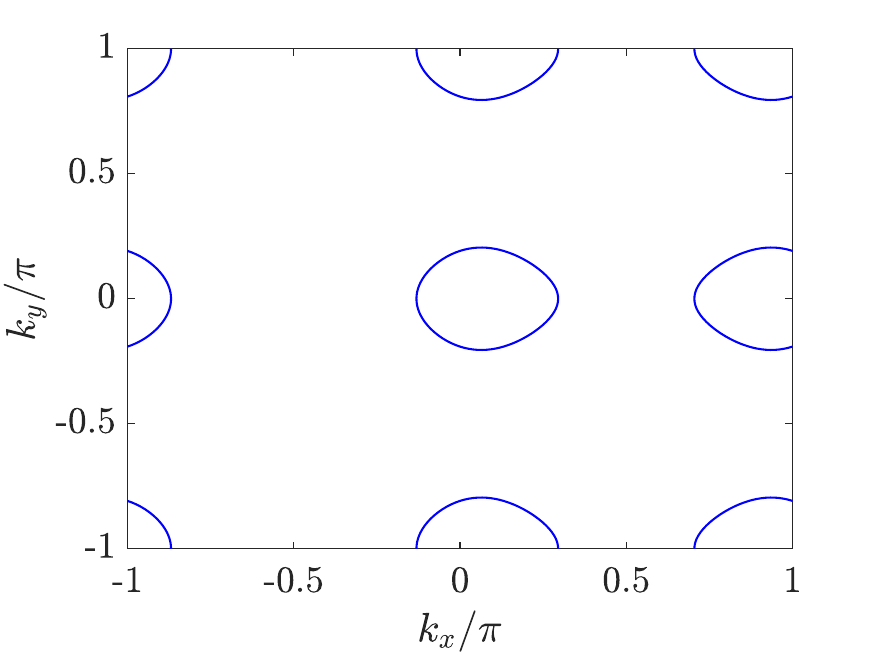}
\caption{Exceptional rings viewed in the $k_{x}-k_{y}$ plane, given by Eq. (%
\protect\ref{RMepring}) with $\protect\alpha= \protect\beta=0.6$. Each loop
represent a pair of exceptional rings corresponding $\sin k_{\text{\textrm{c}%
}z}=\pm \protect\beta$. }
\label{fig2}
\end{figure}

\begin{figure*}[tbh]
\centering
\includegraphics[width=1\textwidth]{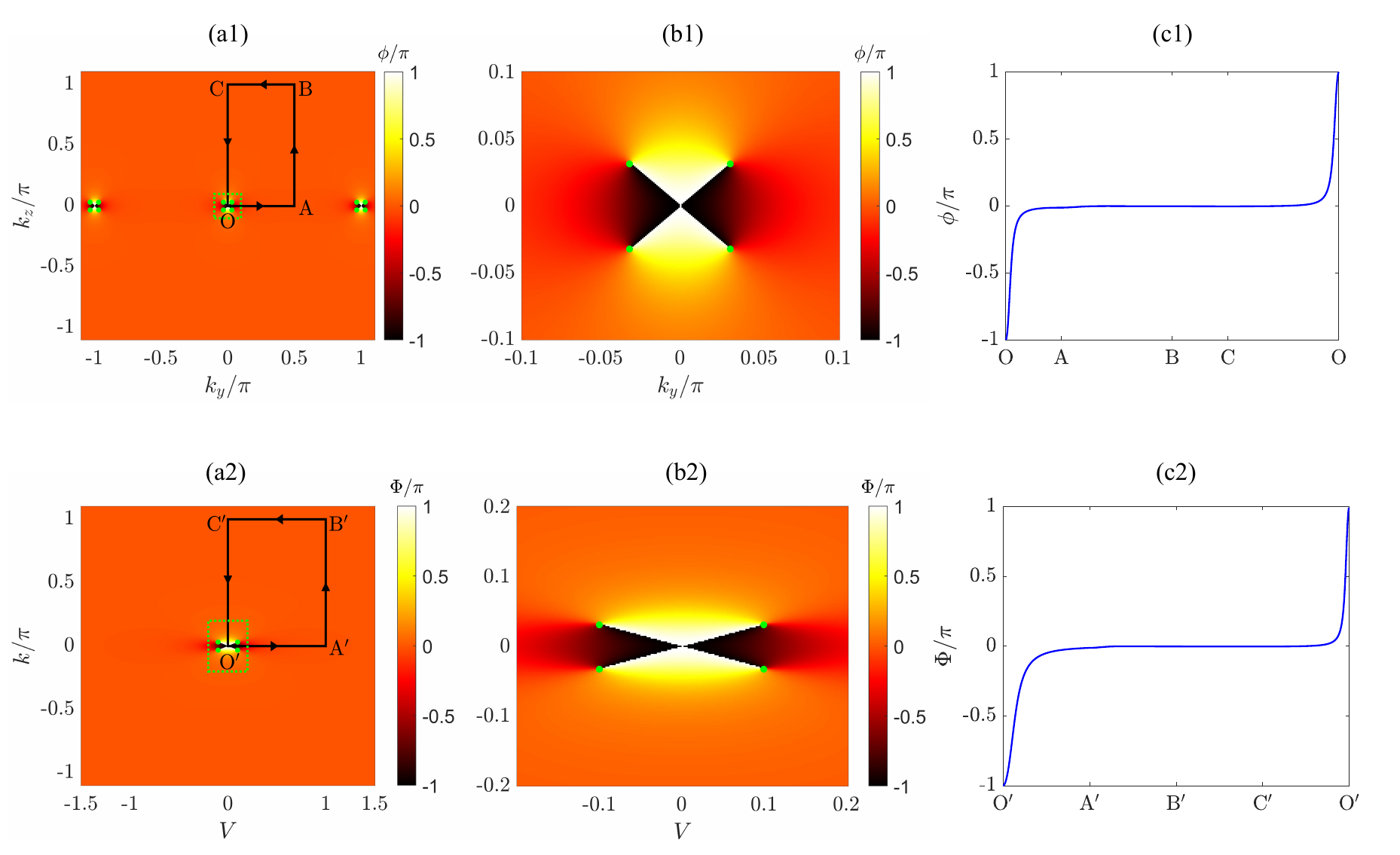}
\caption{Contour plots of $\protect\phi = \protect\phi_{+}+\protect\phi_{-}$
from Eq. (\protect\ref{argument}) ($\Phi = \Phi_{+}+\Phi_{-}$ from Eq. (%
\protect\ref{Phi})) on the $k_{y}-k_{z}$ ($V-k$) plane with $k_{x}=0$ ($%
\protect\delta=0$) and $\protect\alpha=\protect\beta=0.1$. Green dots mark
the exceptional points, symmetric about both axes. O (O$^{\prime }$) denotes
the origin. Black rectangles in (a1) and (a2) encircle single exceptional
points; (c1) and (c2) show the corresponding values along the arrows,
yielding a winding number of 1. Panels (b1) and (b2) magnify the regions
within the green-dashed frames of (a1) and (a2).}
\label{fig3}
\end{figure*}

\section{Exceptional rings as vortex filaments}

\label{Field induced by EP rings}

For the Hermitian case, the topology of the nodal ring is characterized by
invariant quantities, which are abstracted from the set of eigenstates $%
\left\{ \left\vert \psi _{\mu ,\nu }\right\rangle \right\} $\ at $\beta =0$.
Unlike the Hermitian case, the topology of the nodal ring is characterized
by the invariant quantity, which is abstracted from the spectrum $\left\{
\varepsilon _{\mu ,\nu }\right\} $. Based on the previous study in Refs. 
\cite{liu2021higher,kawabata2019}, we introduce a vector field defined as
follows 
\begin{equation}
\mathbf{P=}\nabla _{\mathbf{k}}\left( \phi _{+}+\phi _{-}\right) ,
\label{field}
\end{equation}%
where $\phi _{\pm }$ is the argument of $\varepsilon _{\mu ,\pm }^{2}$,
respectively, i.e., 
\begin{equation}
\varepsilon _{\mu ,\pm }^{2}=\left\vert \varepsilon _{\mu ,\pm
}^{2}\right\vert e^{i\phi _{\pm }}.  \label{argument}
\end{equation}%
In the Appendix, we show that field $\mathbf{P}$ describes a free vortex
flow induced by the exceptional rings acting as vortex filaments. In
general, vortex filaments are idealized, infinitesimally thin lines of
concentrated vorticity, while a free vortex flow describes circular motion
where fluid velocity decreases inversely with radius. This provides a clear
physical picture of the topological structure.

We introduce the winding number to characterize the topology of the EP
rings, which is defined as 
\begin{equation}
w=\oint_{\mathcal{L}}\mathbf{P}\cdot \frac{d\mathbf{k}}{2\pi },  \label{w}
\end{equation}%
where $\mathcal{L}$ denotes an arbitrary closed loop in the momentum space.
Detailed derivations in Appendix A show that the winding number is an
integer with possible values $\left( 0,\pm 1\right) $, determined by the
geometric relationship between the closed loop $\mathcal{L}$ and the two
exceptional rings. The two exceptional rings act as two vortex filaments of
a free vortex flow carrying opposite circulations. When the parameter $\beta 
$\ passes through zero, the two exceptional rings swap places. In this
context, the 3D Hermitian topological semimetal is the boundary separating
two quantum phases identified by two configurations of exceptional rings.
This is consistent with the fact that $\left\vert \psi _{\mu ,\nu }\left( 
\mathbf{k}_{\text{\textrm{c}}}\right) \right\rangle $ is non-analytic at $%
\beta =0$. We demonstrate the contour of $\mathbf{P}$ in Fig. \ref{fig1}
(a2), (b2) and (c2) for several representative configurations of exceptional
rings.

\section{Rice-Mele chain}

\label{Rice-Mele chain}

In the previous sections, we investigated the non-Hermitian semimetal using
the Hamiltonian in momentum space. However, the Hamiltonian $H(\mathbf{k})$
given in Eq. (\ref{Hk})\textbf{\ }is not the periodic function of $\mathbf{k}
$, based on which one cannot obtain a 3D lattice system. On the other hand,
the corresponding 3D lattice system\ probably consists of somewhat
complicated hopping and on-site terms. This would be an obstacle to
fabricating the sample and detecting the topological features in
experiments. In this section, we propose a spinful RM chain model and show
that it has similar topological features in parameter space. We also propose
an experimental scheme to measure the topological index.

\subsection{Hamiltonian and solution}

\label{Hamitlonian and solution}

Consider a 1D RM model on a $2N$ lattice with spin-orbit coupling. The
Hamiltonian is 
\begin{equation}
H=\sum\limits_{\sigma =\uparrow ,\downarrow }H_{\sigma }+H_{\mathrm{so}},
\end{equation}%
where the coupling-free part is 
\begin{eqnarray}
H_{\sigma } &=&\sum\limits_{j=1}^{N}\sum\limits_{\sigma =\uparrow
,\downarrow }\left[ \left( \delta -1\right) c_{2j-1,\sigma }^{\dag
}c_{2j,\sigma }+c_{2j,\sigma }^{\dag }c_{2j+1,\sigma }+\mathrm{h.c.}\right] 
\notag \\
&&-\sum\limits_{j=1}^{2N}\sum\limits_{\sigma =\uparrow ,\downarrow }\left(
-1\right) ^{j+\sigma }Vc_{j,\sigma }^{\dag }c_{j,\sigma }],
\end{eqnarray}%
and the spin-orbit coupling term is 
\begin{equation}
H_{\mathrm{so}}=\lambda \sum\limits_{j=1}^{N}\sum\limits_{\sigma =\uparrow
,\downarrow }\left( c_{2j-1,\sigma }^{\dag }c_{2j,-\sigma }+\text{\textrm{%
h.c.}}\right) .
\end{equation}%
Here, $c_{j,\sigma }^{\dag }$($c_{j,\sigma }$) is the fermion creation
(annihilation) operator at the $j$th site. $H_{\uparrow }$ and $%
H_{\downarrow }$ represent two independent RM models with opposite spin, and
the on-site potential $V$ is spin-dependent with $\left( -1\right)
^{\uparrow }=-\left( -1\right) ^{\downarrow }=1$. The spin-orbit coupling
strength is $\lambda $. It is equivalent to a spinless fermion ladder
system, which is experimentally realizable in waveguide arrays or acoustic
systems \cite{xia2026simple,kang2023topological,chen2025non}. With periodic
boundary conditions, i.e., $c_{2N+1,\sigma }=c_{1,\sigma }$, the Hamiltonian
possesses translational symmetry. Then, employing Fourier transformation as
follows 
\begin{equation}
\left( 
\begin{array}{c}
c_{2j-1,\uparrow } \\ 
c_{2j,\uparrow } \\ 
c_{2j-1,\downarrow } \\ 
c_{2j,\downarrow }%
\end{array}%
\right) =\frac{e^{-ikj}}{\sqrt{N}}\left( 
\begin{array}{c}
a_{k,\uparrow } \\ 
b_{k,\uparrow } \\ 
b_{k,\downarrow } \\ 
a_{k,\downarrow }%
\end{array}%
\right) ,
\end{equation}%
where $k=2n\pi /N$ ($n=0,1,\cdots N-1$), the Hamiltonian can be written as 
\begin{equation}
H=\sum_{2\pi >k\geqslant 0}\Psi _{k}^{\dag }h_{k}\Psi _{k}.
\end{equation}%
Here, we define the operator vector $\Psi _{k}^{\dag }=\left( 
\begin{array}{cccc}
a_{k,\uparrow }^{\dag } & b_{k,\uparrow }^{\dag } & a_{k,\downarrow }^{\dag }
& b_{k,\downarrow }^{\dag }%
\end{array}%
\right) $, and the Hamiltonian matrix takes the form%
\begin{equation}
h_{k}=\left( 
\begin{array}{cccc}
V & \zeta & \lambda & 0 \\ 
\zeta ^{\ast } & -V & 0 & \lambda \\ 
\lambda & 0 & V & \zeta ^{\ast } \\ 
0 & \lambda & \zeta & -V%
\end{array}%
\right) ,
\end{equation}%
with the momentum-dependent factor $\zeta =\delta -1+e^{ik}$.

As with the Hamiltonian in Eq. (\ref{chiral H}), the Bloch Hamiltonian
satisfies 
\begin{equation}
Sh_{k}S^{-1}=-h_{k},  \label{chiral h}
\end{equation}%
where the unitary operator 
\begin{equation}
S=\left( 
\begin{array}{llll}
0 & 0 & 0 & -1 \\ 
0 & 0 & 1 & 0 \\ 
0 & 1 & 0 & 0 \\ 
-1 & 0 & 0 & 0%
\end{array}%
\right) ,
\end{equation}%
implements the chiral symmetry. Consequently, the spectrum is symmetric
about zero. The spectrum of $h_{k}$ consists of four branches

\begin{equation}
\epsilon _{\mu ,\nu }=\mu \sqrt{\left( \lambda +\nu R\right) ^{2}+\sin ^{2}k}%
,  \label{spectrum}
\end{equation}%
with band indices $\mu =\pm $, $\nu =\pm $, and the $k$-dependent factor 
\begin{equation}
R=\sqrt{V^{2}+\left( \delta -1+\cos k\right) ^{2}}.
\end{equation}%
Similarly, with the complex spin-orbit coupling defined as 
\begin{equation}
\lambda =\alpha +i\beta ,
\end{equation}%
the corresponding eigenstates of $h_{k}$ are given by 
\begin{equation}
\left\vert \phi _{\mu ,\nu }\right\rangle =\left( 
\begin{array}{l}
\lambda ^{2}\left[ \zeta ^{\ast }\left( V+\epsilon _{\mu ,\nu }\right)
+\zeta \left( \epsilon _{\mu ,\nu }-V\right) \right] \\ 
\Gamma ^{2}-\lambda ^{2}\left[ \left( V+\epsilon _{\mu ,\nu }\right)
^{2}+\zeta ^{2}\right] \\ 
\lambda \left( \lambda ^{2}\zeta -\zeta ^{\ast }\Gamma \right) \\ 
\lambda \left[ \Gamma \left( V-\epsilon _{\mu ,\nu }\right) -\lambda
^{2}\left( V+\epsilon _{\mu ,\nu }\right) \right]%
\end{array}%
\right) ,
\end{equation}%
with the $k$-dependent factor%
\begin{equation}
\Gamma =\zeta \zeta ^{\ast }+V^{2}-\epsilon _{\mu ,\nu }^{2}.
\end{equation}

Accordingly, the exceptional point condition $\epsilon _{\mu ,\nu }=0$
leads, for $\alpha >0$, to%
\begin{eqnarray}
&&\sqrt{V_{\mathrm{c}}^{2}+\left( \delta _{\mathrm{c}}-1+\cos k_{\mathrm{c}%
}\right) ^{2}}=\alpha ,  \notag \\
&&\sin k_{\mathrm{c}}=\pm \beta .
\end{eqnarray}%
These equations define two circles of radius $\alpha $ with distinct centers
in the ($\delta ,V$) plane, corresponding to different values of $k$.

For $v =-$, the zero-energy eigenstates are given by

\begin{equation}
\left\vert \phi _{\mu ,-}^{\mathrm{c}}\right\rangle =\left( 
\begin{array}{l}
V_{\mathrm{c}}\sin k_{\mathrm{c}}/\beta \\ 
\alpha +\left( \delta -1+\cos k_{\mathrm{c}}\right) \sin k_{\mathrm{c}}/\beta
\\ 
-\left( \alpha \sin k_{\mathrm{c}}/\beta +\delta -1+\cos k_{\mathrm{c}%
}\right) \\ 
V_{\mathrm{c}}%
\end{array}%
\right).
\end{equation}
At the exceptional ring, the two states $\left\vert \phi _{+,-}^{\mathrm{c}%
}\right\rangle $\ and $\left\vert \phi _{-,-}^{\mathrm{c}}\right\rangle $\
coalesce. In the limit $\beta =0$, the condition $\epsilon _{\mu ,\nu }=0$
reduces to 
\begin{eqnarray}
&&\sqrt{V_{\mathrm{c}}^{2}+\left( \delta _{\mathrm{c}}-1+\cos k_{\mathrm{c}%
}\right) ^{2}}=\alpha ,  \notag \\
&&\sin k_{\mathrm{c}}=0,
\end{eqnarray}%
the corresponding zero-energy eigenstates can be expressed as 
\begin{equation}
\left\vert \phi _{+,-}^{\mathrm{c}}\right\rangle =\left( 
\begin{array}{l}
-\zeta _{\mathrm{c}} \\ 
V_{\mathrm{c}} \\ 
0 \\ 
\alpha%
\end{array}%
\right) ,\left\vert \phi _{-,-}^{\mathrm{c}}\right\rangle =\left( 
\begin{array}{l}
-V_{\mathrm{c}} \\ 
-\zeta _{\mathrm{c}} \\ 
\alpha \\ 
0%
\end{array}%
\right),
\end{equation}%
with $\zeta _{\mathrm{c}}=\delta _{\mathrm{c}}-1+\cos k_{\mathrm{c}}$. These
two states are orthogonal
\begin{equation}
\left\langle \phi _{+,-}^{\mathrm{c}}|\phi _{-,-}^{\mathrm{c}}\right\rangle
=0.
\end{equation}%
The exceptional point structure of the RM model resembles that of the
Hamiltonian $H(\mathbf{k})$ discussed above.

\subsection{Connection with 3D model}

\label{Connection with 3D model}

In this subsection, we establish the connection between the RM model and the
matrix $H\left( \mathbf{k}\right) $, given in Eq. (\ref{Hk}), enabling us to
demonstrate the obtained results within the RM model. Under the parameter
transformation%
\begin{equation}
(\delta ,V,k)\rightarrow (\sin k_{x},\sin k_{y},k_{z}),
\label{Ptransformation}
\end{equation}%
the Hamiltonian takes the form%
\begin{equation}
h_{\mathbf{k}}=\left( 
\begin{array}{cccc}
\sin k_{y} & \zeta & \lambda & 0 \\ 
\zeta ^{\ast } & -\sin k_{y} & 0 & \lambda \\ 
\lambda & 0 & \sin k_{y} & \zeta ^{\ast } \\ 
0 & \lambda & \zeta & -\sin k_{y}%
\end{array}%
\right) ,
\end{equation}%
where $\zeta =\sin k_{x}-1+e^{ik_{z}}$, rendering the system periodic in 3D
momentum space. Consequently, the $\phi _{\pm }$ and $\mathbf{P}$ fields are
obtained as periodic functions in the 3D momentum space. The singularities
of these fields constitute exceptional loops satisfying 
\begin{eqnarray}
&&\sqrt{\sin ^{2}k_{\text{\textrm{c}}y}+\left( \sin k_{\text{\textrm{c}}%
x}-1+\cos k_{\text{\textrm{c}}z}\right) ^{2}}=\alpha ,  \notag \\
&&\sin k_{\text{\textrm{c}}z}=\pm \beta .  \label{RMepring}
\end{eqnarray}%
In contrast to the circular exceptional rings of $H\left( \mathbf{k}\right) $%
, those in the RM model exhibit a more complex geometry.\ Fig. \ref{fig1}
shows the field lines of $\mathbf{P}$ for the RM model, compared with those
for $H\left( \mathbf{k}\right) $. In Fig. \ref{fig2} we plot the exceptional
loops for a set of representative parameters in the $k_{x}-k_{y}$ plane,
which appear in pairs at opposite values of $k_{\text{\textrm{c}}z}$. In
Fig. \ref{fig3}(a1) we plot the field $\phi =\phi _{+}+\phi _{-}$ in the $%
k_{y}-k_{z}$ plane with zero $k_{\text{\textrm{c}}x}$.

We can see that $H\left( \mathbf{k}\right) $\ can be regarded as the
effective Hamiltonian of $h_{\mathbf{k}}$\ according to the following
analysis. We note that in the vicinity of $\left( k_{x},k_{y},k_{z}\right)
=\left( 0,0,0\right) $, the Hamiltonian can be approximated as%
\begin{equation}
\bar{h}_{\mathbf{k}}=\left( 
\begin{array}{cccc}
k_{y} & k_{x}+ik_{z} & \lambda & 0 \\ 
k_{x}-ik_{z} & -k_{y} & 0 & \lambda \\ 
\lambda & 0 & k_{y} & k_{x}-ik_{z} \\ 
0 & \lambda & k_{x}+ik_{z} & -k_{y}%
\end{array}%
\right) ,
\end{equation}%
by performing a Taylor series expansion. Furthermore, upon introducing the
unitary transformation 
\begin{equation}
U=\frac{1}{2}\left( 
\begin{array}{cccc}
-i & 1 & i & 1 \\ 
1 & -i & 1 & i \\ 
1 & i & 1 & -i \\ 
i & 1 & -i & 1%
\end{array}%
\right) ,
\end{equation}%
$\bar{h}_{\mathbf{k}}$ is transformed to 
\begin{equation}
\left( 
\begin{array}{llll}
k_{z} & k_{x} & 0 & -k_{y}+\lambda \\ 
k_{x} & -k_{z} & k_{y}+\lambda & 0 \\ 
0 & k_{y}+\lambda & k_{z} & k_{x} \\ 
-k_{y}+\lambda & 0 & k_{x} & -k_{z}%
\end{array}%
\right) ,
\end{equation}%
which coincides with $H\left( \mathbf{k}\right) $ in Eq. (\ref{Hk}) under
the replacement $\lambda \rightarrow m$.

On the other hand, taking the operator vector%
\begin{equation}
\Psi _{\mathbf{k}}^{\dag }=\left( 
\begin{array}{cccc}
a_{\mathbf{k},\uparrow }^{\dag } & b_{\mathbf{k},\uparrow }^{\dag } & a_{%
\mathbf{k},\downarrow }^{\dag } & b_{\mathbf{k},\downarrow }^{\dag }%
\end{array}%
\right) ,
\end{equation}%
the 3D Hamiltonian in real space can be obtained from $\sum_{\mathbf{k}}\Psi
_{\mathbf{k}}^{\dag }h_{\mathbf{k}}\Psi _{\mathbf{k}}$.\ However, the
real-space Hamiltonian describes a 3D lattice system with somewhat
complicated hopping and on-site terms. Therefore, in what follows we
consider the problem using a 1D lattice system.

\subsection{Topology in parameter space}

\label{Topology in parameter space}

Now we turn to the topology of the RM model in the parameter space $(\delta
,V,k)$. Similarly, based on the spectrum in Eq. (\ref{spectrum}), one can
define the arguments of eigenenergies through the relation%
\begin{equation}
\epsilon _{+,\pm }^{2}=\left\vert \epsilon _{+,\pm }^{2}\right\vert e^{i\Phi
_{\pm }},  \label{Phi}
\end{equation}%
which gives 
\begin{equation}
\tan \Phi _{\pm }=\frac{2\beta \left( \alpha \pm R\right) }{\sin
^{2}k+\left( \alpha \pm R\right) ^{2}-\beta ^{2}}.  \label{tPhi}
\end{equation}%
In the case of small $k$, we have%
\begin{eqnarray}
\tan \Phi _{\pm } &\approx &\frac{2\beta \left( \alpha \pm R\right) }{%
k^{2}+\left( \alpha \pm R\right) ^{2}-\beta ^{2}}, \\
R &\approx &\sqrt{V^{2}+\delta ^{2}},
\end{eqnarray}%
which is identical to the expression $\phi _{\pm }$\ in the Appendix by $%
R\rightarrow \rho$ and $k\rightarrow k_{z} $. Therefore, the topological
feature for $H(\mathbf{k})$\ can be simulated via a 1D RM chain in the
parameter space. In Fig. \ref{fig3}(a2) we plot the field $\Phi =\Phi
_{+}+\Phi _{-}$ for comparison with that in 3D momentum space.

From these plots, we note that both fields $\phi $\ and $\Phi $ remain
constant in regions far from the exceptional points (EPs). Consequently, one
can construct a loop containing segments that do not contribute to the path
integral, which simplifies the determination of the winding number. For
instance, we can take a rectangular as a loop in the 3D momentum space $%
\left( k_{x},k_{y},k_{z}\right) $, which contains four segments: (I) $\left(
0,0,0\right) \rightarrow \left( 0,\pi /2,0\right) $; (II) $\left( 0,\pi
/2,0\right) \rightarrow \left( 0,\pi /2,\pi \right) $; (III) $\left( 0,\pi
/2,\pi \right) \rightarrow \left( 0,0,\pi \right) $; (IV) $\left( 0,0,\pi
\right) \rightarrow \left( 0,0,0\right) $. Analytical analysis indicates
that for $\alpha \approx \beta \rightarrow 0$, the path integrals along (II)
and (III) are approximately zero. Then the corresponding winding number is%
\begin{eqnarray}
w &=&\int\nolimits_{0}^{\pi /2}P_{y}\left( 0,k_{y},0\right) \frac{dk_{y}}{%
2\pi }+\int\nolimits_{\pi }^{0}P_{z}\left( 0,0,k_{z}\right) \frac{dk_{z}}{%
2\pi }  \notag \\
&=&\frac{\phi \left( 0,\pi /2,0\right) -\phi \left( 0,0,\pi \right) }{2\pi }%
+1.
\end{eqnarray}%
which can be directly obtained from the spectrum of the 3D lattice $\sum_{%
\mathbf{k}}\Psi _{\mathbf{k}}^{\dag }h_{\mathbf{k}}\Psi _{\mathbf{k}}$.
Alternatively, we can perform a similar procedure for a RM chain, by
replacing $\phi \left( k_{x},k_{y},k_{z}\right) $\ with $\Phi (\delta, V ,k)$%
. We have%
\begin{equation}
w=\frac{\Phi \left( 0,\pi /2,0\right) -\Phi \left( 0,0,\pi \right) }{2\pi }%
+1.
\end{equation}%
In practice, $\Phi \left( \delta, V ,0\right) $\ and $\Phi \left( \delta,
V,\pi \right) $\ can be directly obtained from the band edges. A detailed
explanation is presented in the next section. This renders the experimental
measurement of the winding number feasible. To demonstrate this point, we
perform numerical simulations for the finite RM chain. The results are shown
in Fig. \ref{fig3}(c1) and are consistent with the analytical prediction.

\begin{figure}[tbh]
\centering
\includegraphics[width=0.5\textwidth]{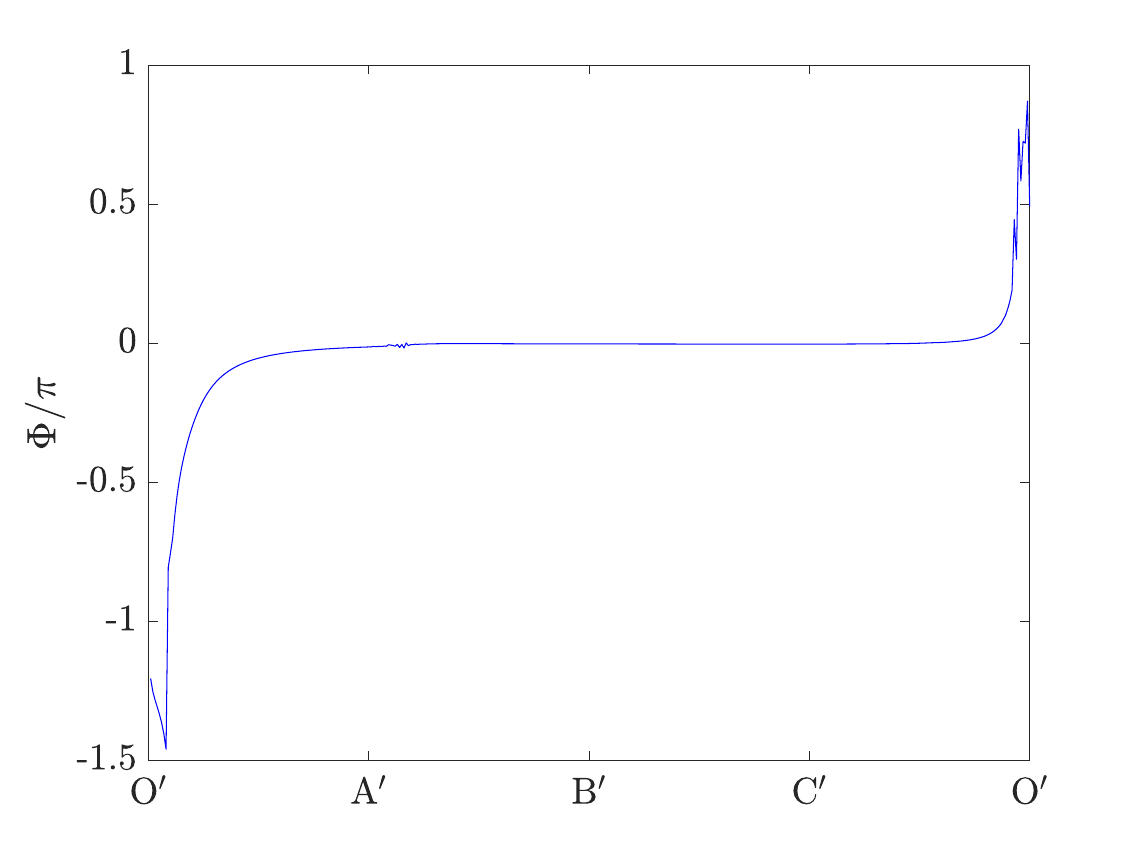}
\caption{Plot of $\Phi$ extracted from the real space energy spectrum of a
ring model along the black rectangle shown in Fig. \protect\ref{fig3}(a2)
with the perturbation described in Eq. (\protect\ref{perturbation}); here $k$
can be interpreted as quasi-momentum. Other parameters are $a=0.1$, $\protect%
\alpha=\protect\beta=0.1$, and $N=200$. }
\label{fig4}
\end{figure}

\section{Robustness to perturbations}

\label{Robustness to perturbations}

In the previous section, we have shown that the winding number can be
obtained from some special energy levels in the spectrum. Moreover, we know
that the boundary conditions and slight perturbations may not affect the
spectrum significantly. Therefore, it is still possible to extract the
winding number from the spectrum in the presence of perturbations. In this
section, we investigate this scheme thoroughly.

We begin with the spectrum of the RM model in the parameter space $(\delta
,V,k)$. From the spectrum in Eq. (\ref{spectrum}), we have%
\begin{eqnarray}
\epsilon _{+,\pm }^{2} &=&\left( \alpha \pm R\right) ^{2}-\beta ^{2}+\sin
^{2}k  \notag \\
&&+2i\beta \left( \alpha \pm R\right) ,
\end{eqnarray}%
and consequently%
\begin{equation}
\text{\textrm{Im}}\left( \epsilon _{+,+}^{2}+\epsilon _{+,-}^{2}\right)
=4\beta \alpha ,
\end{equation}%
which means that the imaginary parts of $\epsilon _{+,\pm }^{2}$ are
symmetric with respect to the constant value $2\beta \alpha $\ (i.e., 
\textrm{Im}$\left( \epsilon _{+,+}^{2}\right) =2\beta \alpha +R$\ and 
\textrm{Im}$\left( \epsilon _{+,+}^{2}\right) =2\beta \alpha -R$). In a
addition, on the $\delta =0$ plane we have 
\begin{equation}
\frac{\partial R}{\partial k}=\frac{\left( 1-\cos k\right) \sin k}{R}%
\geqslant 0
\end{equation}%
for $k\in \left[ 0,\pi \right] $, indicating that \textrm{Im}$\left(
\epsilon _{+,\pm }^{2}\right) $ are monotonic functions of $k$. These
features allow us to distinguish different bands and label the energy level
by $k$ according to the values of \textrm{Im}$\left( \epsilon
_{+,+}^{2}\right) $. We presume that such features remain unchanged in the
presence of perturbations even when the translational symmetry is broken. In
the following, we consider a random perturbation on the on-site potentials,%
\begin{equation}
V_{j}=V+\mathrm{ran}(-a,a)_{j},  \label{perturbation}
\end{equation}%
where $\mathrm{ran}(-a,a)_{j}$ denotes a uniform random number in $(-a,a)$ and $j$ is the unit-cell index with $j\in \left[ 1,N\right] $. A
numerical simulation is performed by taking a set of random numbers $\left\{
V_{j}\right\} $ centered around $V$. Obviously, the perturbation breaks the
translational symmetry even under periodic boundary conditions but preserves
the chiral symmetry. If the perturbation is small enough, it is expected
that the energy levels and their order slightly deviate from those in the
case without perturbation.

Therefore, the winding number can still be obtained from the arguments of
the eigenenergies using the method presented in Fig. \ref{fig3}. To verify
our prediction, we performed numerical simulations on a finite system with
random numbers $\left\{ V_{j}\right\} $. The results are shown in Fig. \ref%
{fig4}. As illustrated, along the same loop as in Fig. \ref{fig3}(a2), the
change in the argument is similar to that in Fig. \ref{fig3}(c2), with the
total change approximately equal to $2\pi $, demonstrating the topological
robustness against perturbations.

\section{Summary}

\label{Summary}

In summary, we have studied the topological features of exceptional nodal
rings emerging in the parameter space of a spinful Rice-Mele chain. The main
contributions are twofold. First, we show that the exceptional rings in a 3D
topological semimetal correspond to a vortex field in momentum space, which
is generated from the complex spectrum. The two exceptional rings act as vortex
filaments of a free vortex flow with opposite circulations. This finding may
be extended to a class of non-Hermitian systems in the future. Second, we
propose an experimental scheme to measure the winding number via a
low-dimensional system. The advantage of this scheme is that only
measurements of single energy levels are needed, rather than those of a
whole energy band.

\acknowledgments This work was supported by the National Natural Science
Foundation of China (under Grant No. 12374461).

\section*{Appendix: Vortex field $\mathbf{P}$}

\label{Appendix}

\appendix\setcounter{equation}{0} \renewcommand{\theequation}{A	%
\arabic{equation}} \setcounter{figure}{0} \renewcommand{\thefigure}{A	%
\arabic{figure}}

\begin{figure}[tbh]
\centering
\includegraphics[width=0.4\textwidth]{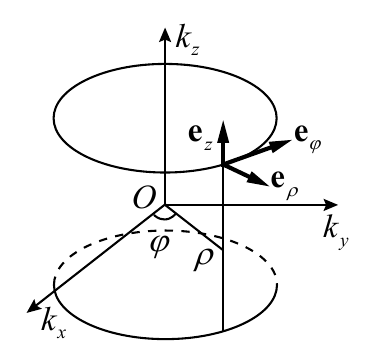}
\caption{Schematic of the cylindrical coordinate system in momentum
space. Here $\protect\rho =\protect\sqrt{k_{x}^{2}+k_{y}^{2}}$ is the radial
distance, $\protect\varphi $ is the azimuthal angle, and $\left( \mathbf{e}_{\protect\rho },\mathbf{e} _{\protect\varphi },\mathbf{e}_{z}\right) $ are the orthonormal basis vectors.}
\label{A1}
\end{figure}

In this appendix, we derive the expression of field $\mathbf{P}$ in Eq. (\ref%
{field}) and demonstrate its features. We note that the function $%
\varepsilon _{\mu ,\nu }\left( \mathbf{k}\right) $ has rotational symmetry
about the $k_{z}$\ axis. It is convenient to take a cylindrical coordinate
system as shown in Fig. \ref{A1}. The unit vectors along the radius $\rho =%
\sqrt{k_{x}^{2}+k_{y}^{2}}$ and the $k_{z}$\ axis\ are $\mathbf{e}_{\rho }$
and $\mathbf{e}_{z}$, respectively. Then the argument given in Eq. (\ref%
{argument}) can be expressed as 
\begin{equation}
\phi _{\pm }=\arctan \frac{2\beta \left( \alpha \pm \rho \right) }{%
k_{z}^{2}+\left( \alpha \pm \rho \right) ^{2}-\beta ^{2}}.
\end{equation}%
Accordingly, the field can be obtained as%
\begin{equation}
\mathbf{P=}\left( \frac{\partial }{\partial \rho }\mathbf{e}_{\rho }+\frac{%
\partial }{\partial k_{z}}\mathbf{e}_{z}\right) \left( \phi _{+}+\phi
_{-}\right) .
\end{equation}%
The explicit expression of the field is%
\begin{equation}
\mathbf{P}=\frac{\mathbf{e}_{1}}{r_{1}}-\frac{\mathbf{e}_{2}}{r_{2}}+\frac{%
\mathbf{e}_{3}}{r_{3}}-\frac{\mathbf{e}_{4}}{r_{4}},
\end{equation}%
where $r_{i}=\left\vert \mathbf{r}_{i}\right\vert $ $\left( i\in \left[ 1,4%
\right] \right) $\ denotes the length of the vector $\mathbf{r}_{i}$,\ and%
\begin{equation}
\mathbf{e}_{i}=\cos \theta _{i}\mathbf{e}_{z}-\sin \theta _{i}\mathbf{e}%
_{\rho },
\end{equation}%
is the tangential unit vector with respect to $\mathbf{r}_{i}/r_{i}$ in the
radial direction.\ Here, the explicit expressions of $\mathbf{r}_{i}$\ are%
\begin{eqnarray}
\mathbf{r}_{1} &=&\left( \rho -\alpha \right) \mathbf{e}_{\rho }+\left(
k_{z}-\beta \right) \mathbf{e}_{z},  \notag \\
\mathbf{r}_{2} &=&\left( \rho -\alpha \right) \mathbf{e}_{\rho }+\left(
k_{z}+\beta \right) \mathbf{e}_{z},  \notag \\
\mathbf{r}_{3} &=&\left( \rho +\alpha \right) \mathbf{e}_{\rho }+\left(
k_{z}+\beta \right) \mathbf{e}_{z},  \notag \\
\mathbf{r}_{4} &=&\left( \rho +\alpha \right) \mathbf{e}_{\rho }+\left(
k_{z}-\beta \right) \mathbf{e}_{z},
\end{eqnarray}%
which are clearly the position vectors of the point $\mathbf{k}$ with
respect to four exceptional points $\alpha \mathbf{e}_{\rho }+\beta \mathbf{e%
}_{z}$, $\alpha \mathbf{e}_{\rho }-\beta \mathbf{e}_{z}$, $-\alpha \mathbf{e}%
_{\rho }-\beta \mathbf{e}_{z}$, and $-\alpha \mathbf{e}_{\rho }+\beta 
\mathbf{e}_{z}$, respectively. In this sense, the physical picture is clear:
the $\mathbf{P}$\ field is the superposition of four free vortex flows
induced by the two exceptional rings acting as vortex filaments. Accordingly, the
winding number, given in Eq. (\ref{w}), is expressed as 
\begin{equation}
w=\frac{1}{2\pi }\oint_{\mathcal{L}}d\left( \theta _{1}-\theta _{2}+\theta
_{3}-\theta _{4}\right) .
\end{equation}%
It accords with the expression of the form 
\begin{eqnarray}
w &=&\oint_{\mathcal{L}}\frac{d\mathbf{k}}{2\pi i}\cdot \nabla _{\mathbf{k}%
}\log \left( \varepsilon _{+,+}^{2}\varepsilon _{+,-}^{2}\right)  \notag \\
&=&\oint_{\mathcal{L}}\frac{d\mathbf{k}}{2\pi i}\cdot \nabla _{\mathbf{k}%
}\log \det \left[ H\left( \mathbf{k}\right) -\varepsilon \left( \mathbf{k}_{%
\text{\textrm{c}}}\right) \right] ,
\end{eqnarray}%
which appeared in the previous work \cite{liu2021higher}.

\input{EPring_8.bbl}

\end{document}

%% file: EPring_8.bbl
%